\documentclass[a4paper]{jpconf}
\usepackage{graphicx}
\usepackage{comment}
\usepackage{amsmath}
\usepackage{amssymb}

\def\bra{\langle }
\def\ket{\rangle }

\begin{document}
\title{Nuclear effects and neutron structure
in
deeply virtual Compton scattering
off $^3$He}

\author{Matteo Rinaldi}

\address{Dipartimento di Fisica e Geologia, 
Universit\`a degli studi di Perugia and
INFN 
sezione di Perugia, Via A. Pascoli 06100
Perugia, Italy}

\ead{matteo.rinaldi@pg.infn.it}

\begin{abstract}
The study of nuclear generalized parton distributions (GPDs) could be a
crucial achievement of hadronic physics since they open new ways to
obtain new information on the structure of bound nucleons, in particular,
to access the neutron GPDs. Here, 
the results of calculations of $^3$He GPDs in Impulse
Approximation 
are
presented.
The calculation of the sum of GPDs  $H+E$, and $\tilde H$, with
the correct limits, will be shown. These quantities, at low momentum
transfer, are largely dominated by the neutron contribution so that $^3$He
is an ideal target for these kind of studies.
Nevertheless the extraction of neutron information from future $^3$He data
could be non trivial.
A procedure, which takes into
account nuclear effects
encoded in IA, is presented. The calculation of $H,E$ and $\tilde H$
allows also to evaluate the cross section asymmetries for deeply virtual
compton scattering at Jefferson Lab kinematics. Thanks to these
observations, DVCS off $^3$He could be an ideal process to access the
neutron
information in the next future.

\end{abstract}

\section{Introduction}
In the last few years, 
generalized parton
distributions (GPDs) 
draw the attention of the 
hadronic physics community since they allow to obtain  new information on
the parton structure of hadrons 
\cite{1l,2l,3l}.
GPDs 
encode the non perturbative 
hadron structure in some peculiar hard exclusive processes. 
In particular, these quantities could give fundamental information on
hadrons, such as the 3-dimensional structure of these systems at
parton level \cite{4l}.
In this  work we focus on the future possibility of shedding
some
light on the so called ``Spin crisis'' problem. 
In principle
this could be realized thanks the relation between the
GPDs and the total angular momentum of partons inside
the hadrons: by subtracting from it the quark helicity contribution to the
hadron spin, it would be possible to access, for the first time, their
orbital angular momentum (OAM) \cite{2l,3l}.

The golden process to study these quantities is the deeply virtual compton
scattering (DVCS), which is
described, at leading twist, mainly by 
$H,~E$ and $\tilde H$ GPDs. This reaction
could be sketched in the following way:
$eH \longmapsto e'H' \gamma$ \ when \ $Q^2\gg M^2$ ($Q^2=-q \cdot q$ \ 
is the momentum transfer
between the leptons beam $e$ and 
$e'$, $\Delta^2$ the one between hadrons $H$ and $H'$ with
momenta $P$ and $P'$, and
$M$ is the nucleon mass).
Another important kinematical variable is the so called
skewedness, $\xi = - \Delta^+/(P^+ + P^{'+})$ 
\footnote{In this paper, $a^{\pm}=(a^0 \pm a^3)/\sqrt{2}$.}.
Even though DVCS is the cleanest process to access the GPDs, the
measurement of these quantities is still a theoretical and experimental
challenge. Despite of these difficulties,  data for proton and nuclear
targets are being analyzed \cite{3c,4c}.  

The study of
hard exclusive processes off nuclear targets, and the relative measurement
of their GPDs, can give new information on possible medium modifications of
the structure of bound nucleons (see  Ref. \cite{5c}). In other words, these
studies are very important 
to have new information on the origin of the so called
EMC effect, an impossible task with the usual DIS studies. To this aim
few-body systems play a special role since, for these targets,  realistic
treatments are possible and exotic effects could be, in
principle, distinguished from the
conventional ones. The use of 
nuclear targets are also necessary, clearly,  to study and to obtain
information on the
neutron GPDs and, in particular, this investigation is the main purpose of
this work.

\section{$^3$He GPDs in impulse approximation}

In order to have a complete flavor decomposition of the GPDs, the proton
data are clearly not sufficient so that the study of the neutron GPDs has
become a priority of the hadron physicists and, as mentioned in the last
section, this could be realized by using nuclear targets. To this aim, among
the light nuclei, $^3$He is an ideal ones due its spin structure, in fact
almost the $90\%$ of the $^3$He spin comes from the neutron one
\cite{6c,7c}. In particular we expect that this nuclear system could be
considered a unique target to extract the neutron $E^n_q$ GPD. In fact
this quantity is, at low momentum transfer,  related to the anomalous
magnetic moments which, for proton and neutron assumes the following values
$k_p \sim1.79~ \mu_N$
and $k_n \sim -1.91~ \mu_N$ (where $\mu_N$ is the nuclear magneton)
respectively so they
are similar in size but with opposite sign. This properties make  isoscalar
nuclei, such as $^2$H and $^4$He, not useful to extract the $E^n_q$  GPD  of
the neutron since the proton contribution basically cancels the
neutron one (see Ref.\cite{5c,26l} for relevant work
on isoscalar light nuclei).
In the $^3$He case, the situation is completely different and to understand
this feature it is sufficient to consider
the dipole magnetic moment of
$^3$He and of the neutron: $\mu_3 \simeq -2.13~\mu_N$ and $\mu_n \simeq
-1.91~
\mu_N$, which
 are very similar and 
would be equal if an independent particle model were valid
for $^3$He. 
Thanks to these properties,
 $^3$He simulates an effective polarized free neutron target so that it
could be used to extract neutron information by properly taking into
account
nuclear effects as in the DIS case, see Ref. \cite{7c}.
From this analysis
 we expect that the sum $\tilde{G}_M^{3,q}(x,\Delta^2,\xi) =
H^{3}_q(x,\Delta^2,\xi)+
E^{3}_q(x,\Delta^2,\xi)$, related to the
contribution of the parton of flavor $q$ to the dipole magnetic moment of
$^3$He, and $\tilde H^3_q(x,\Delta^2, \xi)$, associated to polarized
targets, should be both largely dominated by the
neutron contribution at low momentum transfer. Here it will be just
presented the main passages of the formal
calculation of the $^3$He GPDs.

For a $\frac{1}{2}$ spin target the main quantity which describes the
non perturbative hadron
structure in the analyzed process is the so called light
cone
correlator which is parametrized, at leading twist, by the GPDs
$H_q(x,\Delta^2,\xi)$ and $E_q(x.\Delta^2,\xi)$ in the unpolarized case:

\begin{eqnarray}
 \label{eq1}
&&~~~~~~~~~~~~F_{s's} ^{q,A,\mu} ( x,\Delta^2,\xi) =
\int {\frac{d z^-}{4 \pi}} e^{i x \bar P^+ z^-}
{_A\bra} P' s' | 
\hat O_q^\mu
| P s \ket_A |_{z^+=0,z_\perp = 0} 
=
\nonumber \\
&& \frac{1}{2 \bar P^+} 
\Big [ H_q^A(x,\Delta^2,\xi) \bar u(P',s') 
\gamma^\mu u(P,s) \Big.
+
\Big . 
E_q^A(x,\Delta^2,\xi) \bar u(P',s') 
{i \sigma^{\mu \alpha} \Delta_\alpha \over 2m} u(P,s) \Big ]
\end{eqnarray}

while for the polarized case one has:

\begin{eqnarray}
 \label{eq1t}
&&~~~~~~~~~~~~ \tilde F_{s's} ^{q,A,\mu} ( x,\Delta^2,\xi) =
\int {\frac{d z^-}{4 \pi}} e^{i x \bar P^+ z^-}
{_A\bra} P' s' | 
\tilde O_q^\mu
| P s \ket_A |_{z^+=0,z_\perp = 0} 
=
\nonumber \\
&& \frac{1}{2 \bar P^+} 
\Big [ \tilde H_q^A(x,\Delta^2,\xi) \bar u(P',s') 
\gamma^\mu \gamma^5 u(P,s) \Big.
+
\Big . 
\tilde E_q^A(x,\Delta^2,\xi) \bar u(P',s') 
{i \sigma^{\mu \alpha} \gamma_5 \Delta^+ \over 2m} u(P,s) \Big ]
\end{eqnarray}

where the initial (final)
momentum and helicity are $P(P')$ and $s(s')$, 
respectively, and $\hat O_q^\mu=
\bar \psi_q 
\left(- {z \over 2 } \right)
\gamma^\mu \, 
\psi_q \left( {z \over 2 } \right)
$, $\tilde O_q^\mu=
\bar \psi_q 
\left(- {z \over 2 } \right)
\gamma^\mu \gamma_5 \, 
\psi_q \left( {z \over 2 } \right)
$,
$\bar P=(P+P')/2$,
$\psi_q$ is the quark field, $m$ is the hadron mass.

Since we have to rely on a
$^3$He wave
function, it is necessary to perform a non relativistic limit of the light
cone correlator in order to have a complete consistency. Thanks to this
procedure, simple relations between the GPDs and spin components of the
correlator were found (see Ref.\cite{8c} for details):
 
\begin{figure*}[t]
\vspace{6.5cm}
\includegraphics{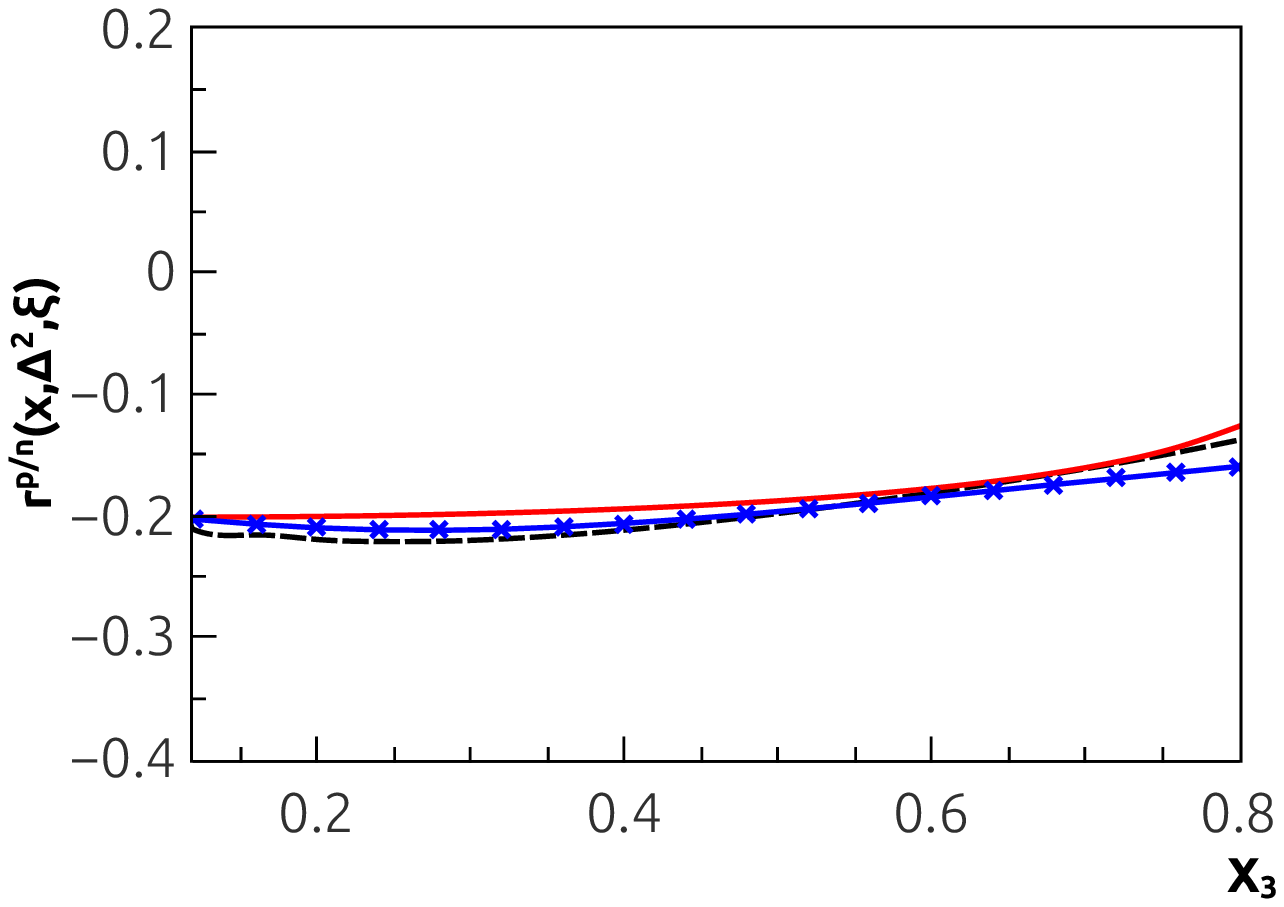}
\includegraphics{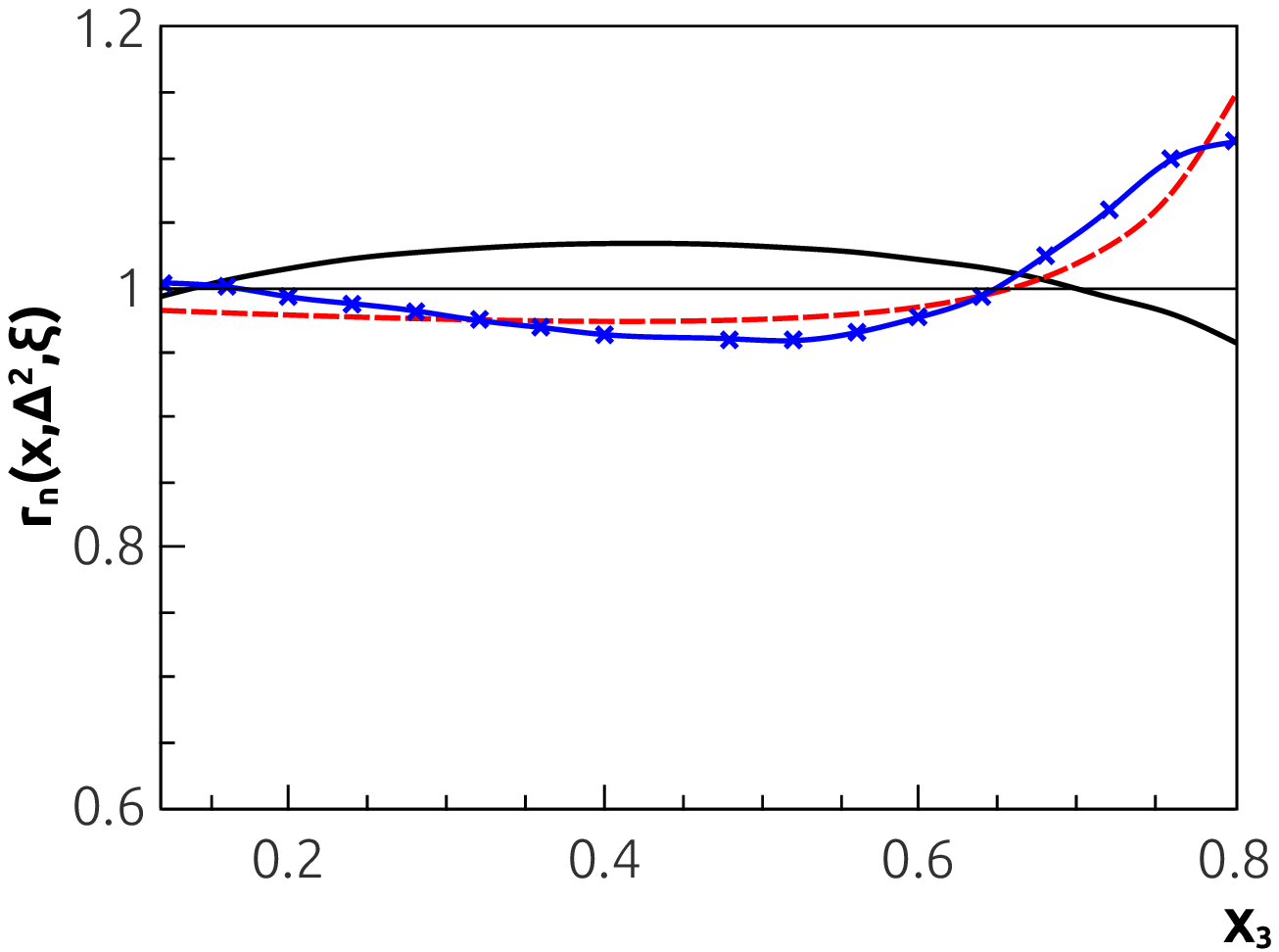}
%
\caption{ 
The ratio $r^{p/n}(x,\Delta^2,\xi)$ and $r_{n}(x,\Delta^2,\xi)$ evaluated
in this region: $-\Delta^2 = 0.1$ GeV$^2$ and $\xi = 0$, where here  $x_3 =
(M_3/M) x$ and using the models of Refs.\cite{13c,37l,38l}
}
\label{fig-1} 
\end{figure*}

\begin{eqnarray}
\label{relazioni}
H_q^A(x,\Delta^2,\xi) &=& \dfrac{\overline{P}^+}{m} F_{++}^{q,A,0}
(x,\Delta^2,\xi)~,
\nonumber \\
\tilde G_M^{A,q}(x,\Delta^2,\xi) &=& \dfrac{2\overline{P}^+}{\Delta_z}
F_{+-}^{q,A,1}(x,\Delta^2,\xi)~.
\nonumber \\
\tilde H_q^A(x,\Delta^2,\xi) &=& \dfrac{\overline{P}^+}{m}~i~
\tilde F_{+-}^{q,A,2}
(x,\Delta^2,\xi)
\label{nr}
\end{eqnarray}
It is remarkable that one of the quantities we are analyzing is the
sum  $\tilde{G}_M^{3,q}(x,\Delta^2,\xi) =
H^{3}_q(x,\Delta^2,\xi)+
E^{3}_q(x,\Delta^2,\xi)$ which is also, in addition to the
connection with the magnetic form factor (ff) of the hadron,  a
fundamental quantity thanks to its relation with the total angular momentum
of partons inside the hadrons (Ji's sum rule \cite{3l}):

\begin{eqnarray}
 J_q^A = \int_{-1}^{1} dx ~x~ \tilde G^{A,q}_M(x,0,0)
\end{eqnarray}
This result justifies our interest in the calculation of this particular
combination of the GPDs. Starting from the NR 
relations, one can describe the light cone correlator in terms of quantities
which depend on $^3$He wave function, by properly applying the IA (see
Ref.\cite{9c} for details). Using then the same GPDs relations,
Eqs.(\ref{nr}), for the free nucleonic contributions, one finds:

\begin{eqnarray}
 \tilde G_M^{3,q}(x,\Delta^2,\xi)  = 
\sum_N
\int dE 
\int d\vec{p}~
{\left [ P^N_{+-,+-}
-
P^N_{+-,-+} \right](\vec p,\vec p\,',E) }
{\xi' \over \xi}
\tilde G_M^{N,q}(x',\Delta^2,\xi')
\label{new}
\end{eqnarray}
\vskip-3mm
\noindent
and
\begin{eqnarray}
{\tilde H_{q}^3(x,\Delta^2,\xi)}  = 
\sum_N
\int dE 
\int d\vec{p}
\,
{\left [ P^N_{++,++}
-
P^N_{++,--} \right](\vec p,\vec p\,',E) }
{\xi' \over \xi}
{\tilde H^{N}_q(x',\Delta^2,\xi')}~,
\label{new1}
\end{eqnarray}

where here  $x'$ and $\xi'$, the variables 
for the bound nucleon 
GPDs, have been introduced, with $p \, (p'= p + \Delta)$
its 4-momentum in the initial (final) state and 
proper components appear of the spin dependent,
one body off diagonal spectral function:
\begin{eqnarray}
 \label{spectral1}
 P^N_{SS',ss'}(\vec{p},\vec{p}\,',E) 
= 
\dfrac{1}{(2 \pi)^6} 
\dfrac{M\sqrt{ME}}{2} 
\int d\Omega _t
\sum_{\substack{s_t}} \langle\vec{P'}S' | 
\vec{p}\,' s',\vec{t}s_t\rangle_N
\langle \vec{p}s,\vec{t}s_t|\vec{P}S\rangle_N~.
\end{eqnarray}
Here $E= E_{min} +E_R^*$, 
where $E^*_R$ is the excitation energy 
of the full interacting two-body recoiling system.
The main ingredient appearing in the definition
Eq. (\ref{spectral1}) is
the intrinsic overlap integral
\begin{equation}
\langle \vec{p} ~s,\vec{t} ~s_t|\vec{P}S\rangle_N
=
\int d \vec{y} \, e^{i \vec{p} \cdot \vec{y}}
\langle \chi^{s}_N,
\Psi_t^{s_t}(\vec{x}) | \Psi_3^S(\vec{x}, \vec{y})
 \rangle~
\label{trueover}
\end{equation}

\begin{figure*}[t]
\vspace{6.5cm}
\includegraphics{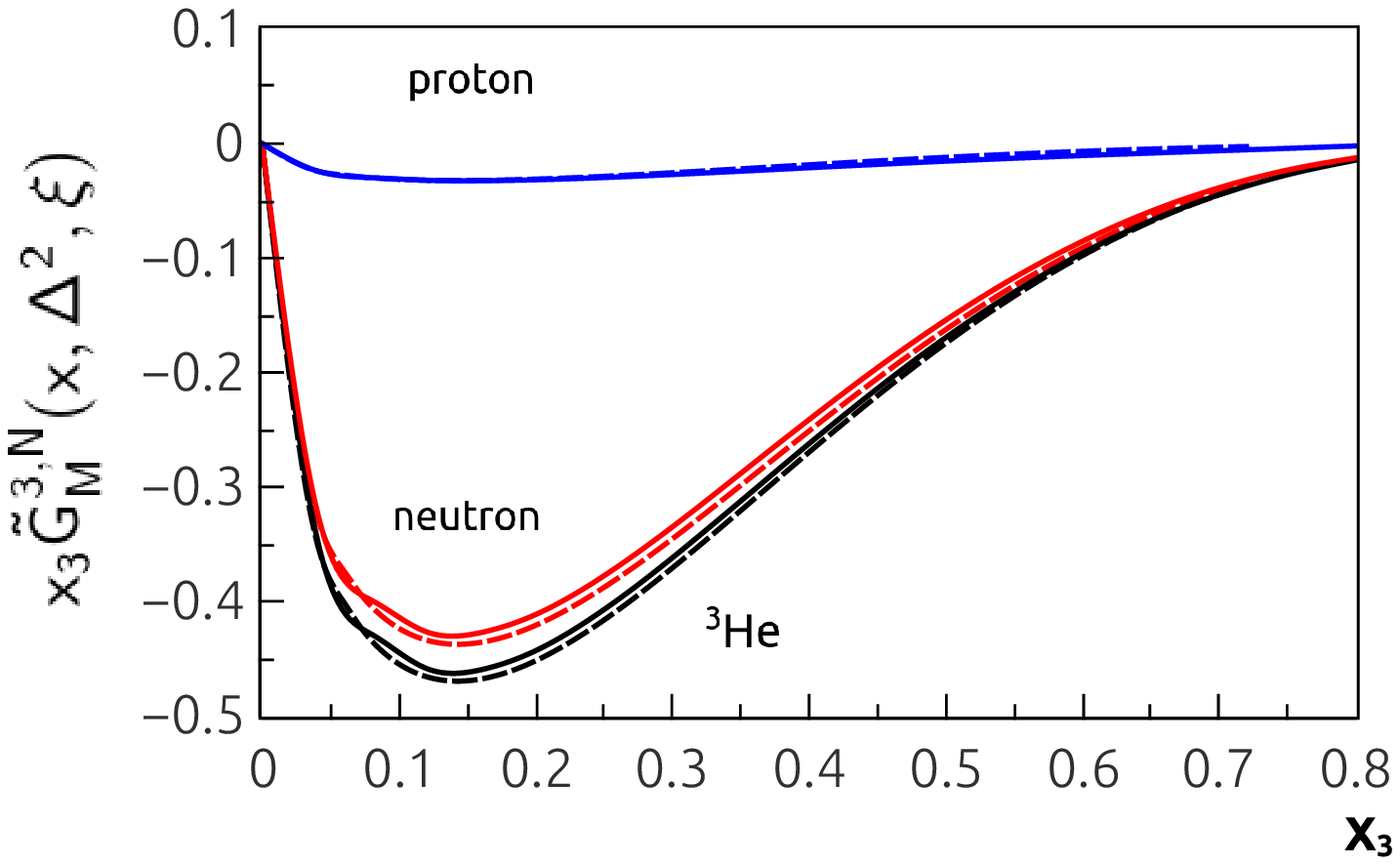}
\includegraphics{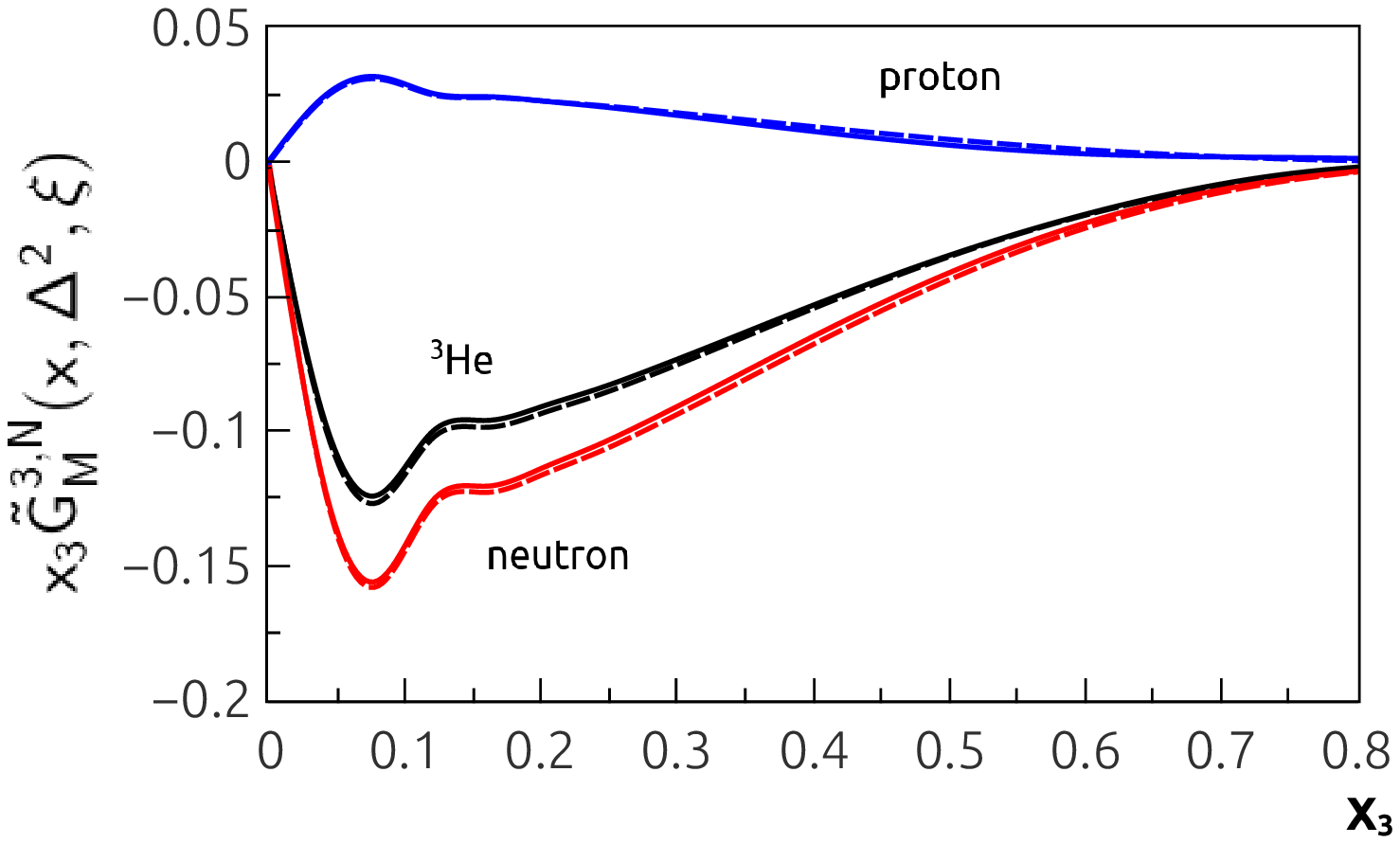}
%
\caption{ 
The  GPD $x_3 \tilde{G}^{3}_M(x,\Delta^2,\xi)$, 
where $x_3 = (M_3/M) x$
and $\xi_3 = (M_3/M) \xi$, shown in the forward limit (left panel)
and at $\Delta^2 = -0.1
\ \mbox{GeV}^2$ and $\xi_3=0.1$ (right panel), 
together with the neutron and the proton
contribution. Solid lines represent the full IA result, Eq. (\ref{new}),
while the dashed ones correspond to the approximation
Eq. (\ref{sgug}).} 
\end{figure*}

between the $^3$He wave function,
$\Psi_3^S$,  
and the final state, described by two wave functions: 
{\sl i)}
the
eigenfunction $\Psi_t^{s_t}$, with eigenvalue
$E = E_{min}+E_R^*$, of the state $s_t$ of the intrinsic
Hamiltonian describing the system of two {\sl interacting}
nucleons with relative momentum $\vec{t}$, 
which can be either
a bound 
or a scattering state, and 
{\sl ii)}
the plane wave representing 
the free nucleon $N$ in IA.
For a numerical evaluation of Eqs. (1) and (2),
the overlaps, Eq. (4), appearing in Eq. (3)
and corresponding to the analysis presented in Ref.
\cite{10c} in terms of 
AV18  \cite{11c} wave functions
\cite{12c}, 
have been used. Clearly the accuracy  of this calculation, since the NR
relativistic spectral function has been used, will be of the order 
${\cal{O}} 
\left ( {\vec p\,^2 / M^2},{\vec \Delta^2 / M^2} \right )$.

Since no data for $^3$He GPDs are actually available, the only possible
checks for our calculations will be the GPDs properties, in particular the
forward limit and the first moment. In Ref.\cite{9c} it
has been illustrated that
$H^3_q(x,\ \Delta^2,\xi)$ fulfills these constrains while,  in
Ref.\cite{8c},
the quantity $
\sum_q \int dx \, \tilde G_M^{3,q}(x,\Delta^2,\xi) = G_M^3(\Delta^2)
$, which should give the magnetic ff of $^3$He, has
been calculated, since no forward limit is defined for the
$E_q(x,\Delta^2,\xi)$ GPD. Our result is in agreement with
 the one-body part of the
AV18 calculation,
presented in Ref.\cite{14c}, and also for 
$-\Delta^2 \le 0.15$ GeV$^2$, which is the relevant kinematics condition for
the coherent DVCS at JLab, our results compare  well with data.
Concerning $\tilde H_q^3(x,\Delta^2,\xi) $, its first moment is the axial
ff of $^3$He but, since this quantity is poorly known, the only possible
check for this GPD is the forward limit, which is correctly recovered. 

In order to present the main results for the calculations of $^3$He GPDs and
estimating the proton and neutron contribution, describing the IA, it is
necessary to explain some details on the used model for the free nucleonic
GPDs, necessary to perform  numerical evaluations.

\begin{figure*}[t]
\vspace{6.5cm}
\includegraphics{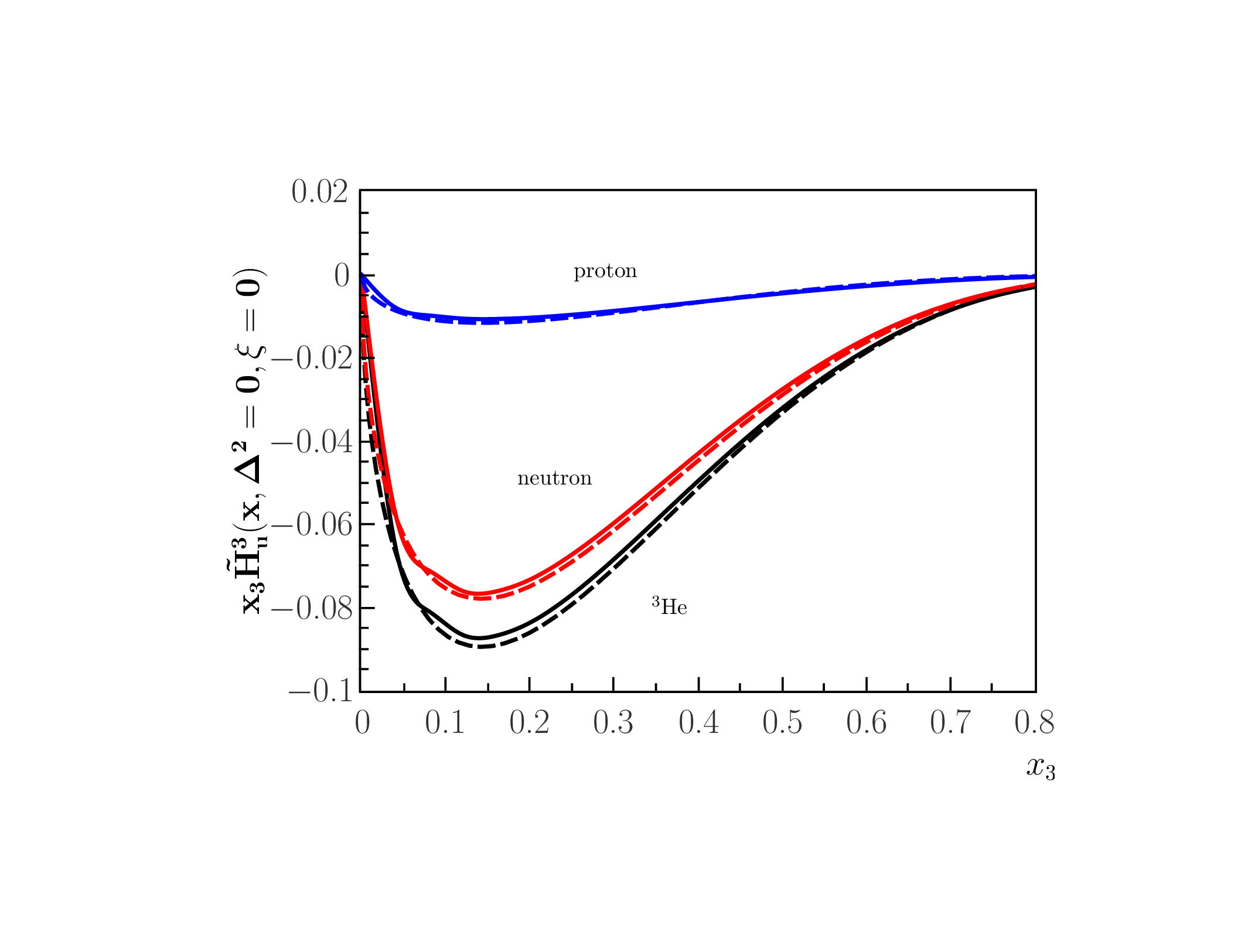}
\includegraphics{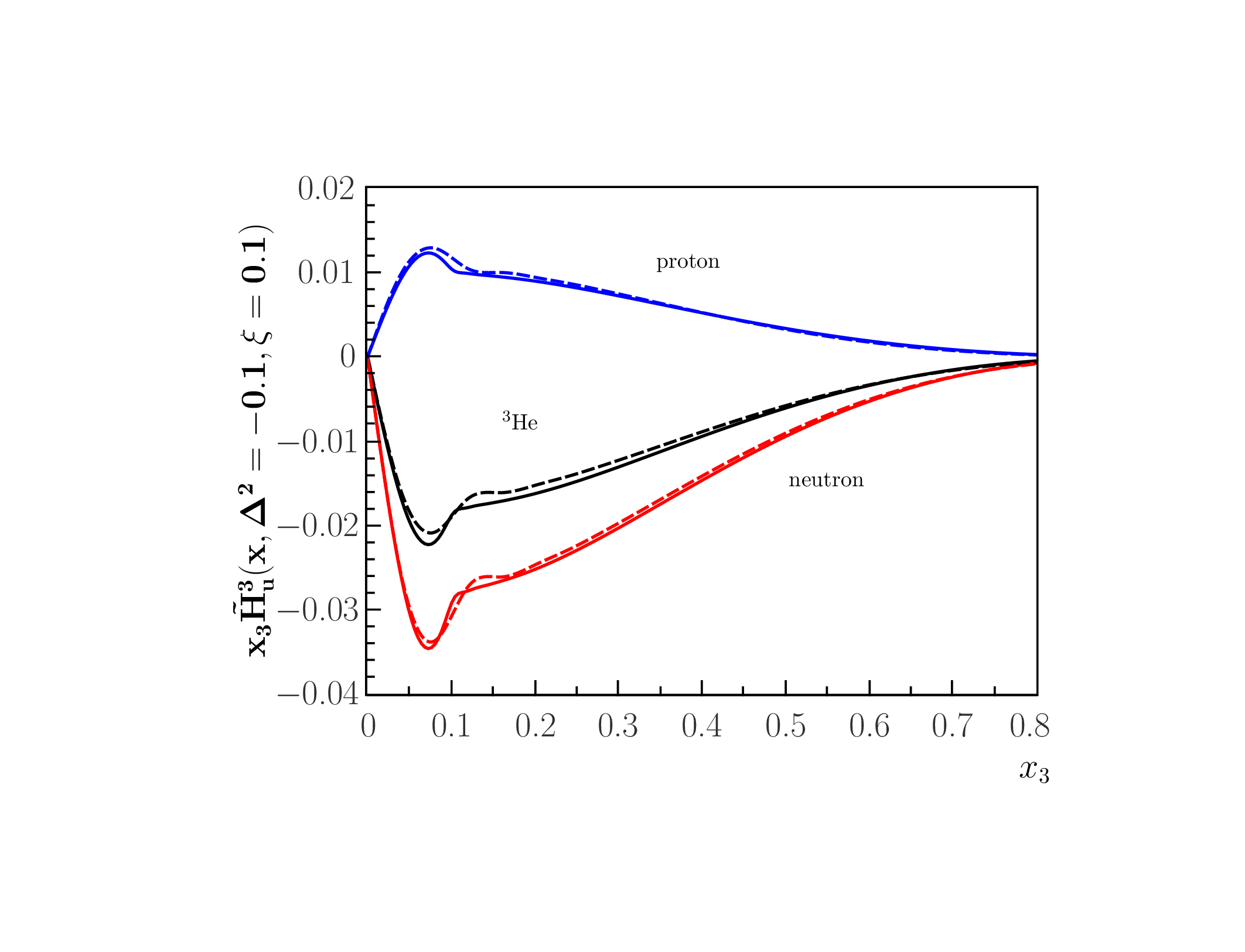}
%
\caption{ 
The  GPD $x_3 \tilde{H}^{3}_u(x,\Delta^2,\xi)$ for the flavor $q=u$  shown
in the forward limit (left panel)
and at $\Delta^2 = -0.1
\ \mbox{GeV}^2$ and $\xi_3=0.1$ (right panel), 
together with the neutron and the proton
contribution. Solid lines represent the full IA result, Eq. (\ref{new1}),
while the dashed ones correspond to the approximation
Eq. (\ref{sguh}).}
\label{fig-2} 
\end{figure*}

\section{Nucleonic models of GPDs}
In this work, as pointed out in the previous section, we need to perform a
 numerical evaluation of $^3$He GPDs; in particular, as
suggested by the behavior of the $^3$He wave function, these quantities
should be  dominated by the neutron contribution. This feature 
is provided by
nuclear properties and should not  be inferred from
the nucleonic structure so that, for the calculation of $\tilde
G_M^3(x,\Delta^2,\xi)$, we have used three completely different models:

\begin{itemize}
 \item  for our first calculation we have used a very  simple model,
Ref.\cite{13c}, which  fulfills all the
 properties of GPDs. In this particular case the $\Delta^2$ dependence
is factorized  out from the $x$ and $\xi$ one by introducing the 
contribution of the quark of
flavor $q$ to the nucleonic ff: $F_q(\Delta^2)$. To this aim
experimental values of $F_1^{n,p}(\Delta^2)$ Dirac ff have been used 
\cite{35l}
and  a
parametrization has been chosen in order to
have
a flavor decomposition of these quantities 

This model has been
properly extended to describe the $E_q$ GPD (see Refs.\cite{8c} for
details);

\item we have also used a description of the nucleonic structure  based on
a constituent quark model (along the
lines of Ref.\cite{37l}),
valid therefore in the valance quarks region. This model is 
very different from the previous one, and it does not
assume any kind of
factorization;

\item
for further analysis,  a simple version of the MIT model, Ref.\cite{38l},
has been used. This is a very different scenario since, here, free
relativistic confined quarks are described.

\end{itemize}

It is important to remark that our purpose is to estimate and to extract
neutron information from $^3$He GPDs so our main interest is on the
nucleonic contribution to these quantities, a feature rather independent on
the nucleonic model
chosen, as it will be
shown in the next section. Therefore, for the moment being, for the
$\tilde H_q$ calculation,
we have just extended the first model described above, where a factorization
dependence on $\Delta^2$ is assumed, by properly introducing 
a simple ansatz for
the axial
nucleonic ff, $\tilde F_q(\Delta^2) = \dfrac{1}{\left(1-
\frac{\Delta^2}{M_A^2}\right)^2}$ where $M_A \simeq 5.076~~fm^{-1}$ 
(see Ref.\cite{htilde1} for details).

Clearly, if experiments were planned, more realistic models for nucleonic
GPDs will be adopted and included into the scheme.

\section{Nucleonic contributions to $^3$He GPDs}

Thanks to the nuclear and the nucleonic model described in the previous
sections, and the comfort of all positive checks, the $^3$He GPDs:  $\tilde
G^3_M(x,\Delta^2,\xi) = \underset{q}{\sum}~\tilde G^{3,q}_M(x,\Delta^2,\xi)$
and $\tilde H^3_q(x,\Delta^2,\xi)$ will be presented. Concerning these
quantities, we found that the neutron contribution largely dominates the
$^3$He GPDs, at low momentum transfer, but, increasing $\Delta^2$, the
proton contribution grows up, in particular for $u$ flavor. For the quantity
$\tilde G^3_M(x,\Delta^2,\xi)$, this peculiar behavior is governed by the
magnetic ff of $^3$He, where the proton contribution is basically negligible
for low $\Delta^2$ but, increasing the momentum transfer, the size of the
neutron
contribution decreases (see Ref.\cite{8c}).
For this reason, as already pointed out, since our interest is on the
nucleonic
contributions to $^3$He GPDs,  we have analyzed the ratio
$r^{p/n}(x,\Delta^2,\xi) = \tilde G_M^{3,p}(x,\Delta^2,\xi)/
\tilde G_M^{3,n}(x,\Delta^2,\xi)~$ using the three different models
described in the previous section. The result of this calculation shows and
demonstrates that this quantity, and the relative nucleonic contributions
to $^3$He GPDs, clearly depend on the nuclear effects encoded in the $^3$He
wave
function rather then  the free nucleonic structure. 
In closing this section, let us remind that our calculations show how  the
neutron contribution  largely dominates the $^3$He GPDs, both in the case of
$\tilde G^3_M$ and $\tilde H^3$ in the kinematics region of $-\Delta^2 
\lesssim 0.1$ GeV$^2$, in particular for the $d$ flavor case. Despite the
limited validity of the region of  IA and the neutron dominance, these
studies could be a prerequisite for future experiments since the Ji's sum
rule, which is the main information we are interested in, is only valid  in
the
forward limit, so that it would be
crucial to estimate GPDs at very low momentum transfer.

In spite of these promising results, a
procedure to extract the neutron GPDs, from future $^3$He data, at
different values of momentum transfer, is necessary since the relations
found for the  nuclear GPDs, Eqs.(\ref{new},\ref{new1}), are 
convolution-like equations
involving a nuclear term, the off diagonal spectral function, and a
nucleonic one, which is the free nucleonic GPD. In this scenario the
extraction of the neutron could be not trivial.

\section{Extraction procedure of neutron GPDs from $^3$He}

In this section 
the extraction procedure,
will be presented, together with numerical results,
which allows to obtain the neutron GPDs from future
$^3$He data. To this aim it is sufficient to write the $^3$He GPDs in terms of
the ``light cone spin dependent off-forward momentum distributions'',
$h(g)^3_N(z,\Delta^2,\xi)$ which are peaked around $z=1$ close to the
forward limit.

\begin{eqnarray}
\tilde H^{3}_q(x,\Delta^2,\xi) =   
\sum_N \int_{x_3}^{M_A \over M} { dz \over z}
h_N^3(z, \Delta^2, \xi ) ~
\tilde H^{N}_q \left( {x \over z},
\Delta^2,
{\xi \over z},
\right)~,
\end{eqnarray}

\begin{eqnarray}
\tilde G_M^{3,q}(x,\Delta^2,\xi) =   
\sum_N \int_{x_3}^{M_A \over M} { dz \over z}
g_N^3(z, \Delta^2, \xi ) 
~\tilde G_M^{N,q} \left( {x \over z},
\Delta^2,
{\xi \over z},
\right)~.
\end{eqnarray}

Therefore, in the region delimited by $x_3 \leq 0.7$, one can approximate
the full calculation of $^3$He GPDs with these new and simpler relations:

\begin{eqnarray}
\tilde H^{3}_q(x,\Delta^2,\xi) 
& \simeq &   
\sum_N 
\tilde H^{N}_q \left( x, \Delta^2, {\xi } \right)
\int_0^{M_A \over M} { dz }~
h_N^3(z, \Delta^2, \xi ) 
\nonumber
\\
& = &
G^{3,p,point}_{A}(\Delta^2) 
\tilde H^{p}_q(x, \Delta^2,\xi) 
+ 
G^{3,n,point}_{A}(\Delta^2) 
\tilde H^{n}_q(x,\Delta^2,\xi)~,
\label{sguh}
\end{eqnarray}

\begin{eqnarray}
\tilde G_M^{3,q}(x,\Delta^2,\xi) 
& \simeq &   
\sum_N 
\tilde G_M^{N,q} \left( x, \Delta^2, {\xi } \right)
\int_0^{M_A \over M} { dz }~
g_N^3(z, \Delta^2, \xi ) 
\nonumber
\\
& = &
G^{3,p,point}_{M}(\Delta^2) 
\tilde G_M^{p,q}(x, \Delta^2,\xi) 
+ 
G^{3,n,point}_{M}(\Delta^2) 
\tilde G_M^{n,q}(x,\Delta^2,\xi)~,
\label{sgug}
\end{eqnarray}

where here the axial (A) and magnetic (M) point like ffs have been defined:
 $G^{3,N,point}_{A(M)}(\Delta^2)=\int_0^{M_A \over M} dz \,
h(g)_N^3(z,\Delta^2,\xi).
$ 

These quantities would give the nucleon $N=n,p$ contribution to the nuclear
axial (magnetic) ff if protons and neutrons were 
point-like
particles.

In the magnetic case, in the kinematics region under scrutiny here, we found
that these quantities are  rather independent
on  nuclear potential (see
Ref.\cite{8c} for details) while, concerning the axial case, we found that
these ffs, in the forward limit, reproduce the so called ``effective
polarizations'' of protons ($p_p$) and neutron ($p_n$), whose values, using
the AV18 potential, are: $p_n=0.878$ and $p_p= -0.023$. In this limit
our results reproduce formally and numerically  the formalism obtained in
Ref.\cite{7c} for the polarized DIS off $^3$He. Starting from
Eqs.(\ref{sguh},\ref{sgug}) it is possible to extract the free neutron GPDs
from the $^3$He ones.

\vskip-5mm
\begin{eqnarray}
\tilde H^{n,extr}(x, \Delta^2,\xi)  \simeq  
{1 \over G^{3,n,point}_{A}(\Delta^2)} 
\left\{ \tilde H^3(x, \Delta^2,\xi) 
 -  
G^{3,p,point}_{A}(\Delta^2) 
\tilde H^p(x, \Delta^2,\xi) \right\}~,
\label{extrh}
\end{eqnarray}

\begin{eqnarray}
\tilde G_M^{n,extr}(x, \Delta^2,\xi)  \simeq  
{1 \over G^{3,n,point}_{M}(\Delta^2)} 
\left\{ \tilde G_M^3(x, \Delta^2,\xi) 
 -  
G^{3,p,point}_{M}(\Delta^2) 
\tilde G_M^p(x, \Delta^2,\xi) \right\}~.
\label{extrg}
\end{eqnarray}

In order to test these new relations, we have compared
Eqs.(\ref{sguh},\ref{sgug}) with the full calculation of $^3$ He GPDs, 
Eqs.(\ref{new},\ref{new1}) respectively. The results show that the new
equations, in the kinematical region where the IA is valid, approximate
very well the full calculation for $x \leq 0.7$, where the DVCS can be
experimentally analyzed at the JLab.  In particular it is remarkable that
the only theoretical ingredients for these calculations are the axial
(magnetic)
ffs which are under good theoretical control.
With the comfort of this successful results, in order to simulate the
extraction procedure, since no data for the $^3$He GPDs exist, we have
compared the neutron GPDs extracted, from the approximated relations
Eqs.(\ref{extrh}, \ref{extrg}), with the free neutron GPDs used as input
for the full calculation and using the same model for the proton GPDs. The
results of this comparisons show that this procedure works perfectly for $x
\leq 0.7$.

In order to have more accuracy of this analysis, we have studied the ratio
$r_n(x,\Delta^2,\xi) = \tilde G^{n,extr}_M(x,\Delta^2,\xi)/\tilde
G^n_M(x,\Delta^2,\xi)$ in a region which is beyond the forward limit and
 we have evaluated it using three completely different nucleonic models
\cite{13c,37l,38l}. This result shows that our extraction procedure is
weekly dependent even on the free nucleonic model .

\section{Conclusion}

In closing, we have shown the $^3$He GPDs, which are dominated, at low
momentum transfer, by neutron contribution and, due to the complicated
relation between the nuclear and the nucleonic GPDs, we have proposed an
extraction procedure of the neutron GPDs, which is
able to take into account the nuclear effects encoded in the IA and which is
rather
independent on the free nucleonic model and nuclear potential. These
successful results make  the coherent DVCS off $^3$He  an ideal process to
access the neutron GPDs. 
If higher momentum transfer region will be experimentally accessed, our
work could be implemented by including a relativistic treatment, such as
the on in Ref. \cite{15c}
and/or many body currents,  going beyond the IA.

 Now it will be possible
to study the cross section asymmetries for the DVCS processes the JLab
kinematics since,  for a $\frac{1}{2}$ spin target (see Ref.\cite{16c}), at
leading twist, the three GPDs $H,E$ and $\tilde H$ will mainly contribute.
We have therefore at hand 
all the ingredients to perform this completely new analysis.

I thank Sergio Scopetta for a nice and fruitful collaboration
on this subject in the last three years.
This work was supported in part by the Research Infrastructure
Integrating Activity Study of Strongly Interacting Matter (acronym
HadronPhysic3, Grant Agreement n. 283286) under the Seventh Framework
Programme of the European Community.

\end{document}